\documentclass{aa}
\input{psfig.sty}
\def\simgt{\lower.5ex\hbox{$\;\buildrel>\over\sim\;$}}
\def\simlt{\lower.5ex\hbox{$\;\buildrel<\over\sim\;$}}
\begin{document}

   \thesaurus{03     (11.03.2; 
                      11.04.2; 
                      11.09.1 Tol~35 
                      11.09.1 Tol~3  
                      11.09.1 UM~462 
                      11.19.3)} 
   \title{Near-infrared properties of Blue Compact Dwarf Galaxies: the link 
between solar and low metallicity 
\thanks{Based on observations obtained at the ESO-NTT in La Silla}}


   \author{L. Vanzi\inst{1}, L.K. Hunt\inst{2}, T.X. Thuan\inst{3}}

   \offprints{L. Vanzi}

   \institute{ESO, Alonso de Cordova 3107, Santiago - CHILE\\
              email: lvanzi@eso.org
          \and
              IRA/CAISMI - CNR,
              Largo E. Fermi 5, 50125 Firenze - ITALY
          \and
              Astronomy Department, University of Virginia,
              Charottesville, VA 22903 - U.S.A.
             }

   \date{Received ....; accepted ....}

\authorrunning{Vanzi et al.}
\titlerunning{NIR properties of BCD galaxies}

   \maketitle

   \begin{abstract}
We have obtained near-infrared images and spectra of three blue compact dwarf galaxies 
of intermediate sub-solar metallicity Tol~35, Tol~3 and UM~462. 
This work is part of a larger project aimed to study the star formation and the stellar 
populations of low metallicity galaxies in the near-infrared. 
In this frame work galaxies of intermediate metallicity represent an important 
step in understanding the most extreme cases filling the gap between solar 
and very low metallicity galaxies. 
We have observed HII region like spectra in all three galaxies; in all cases the
star formation episodes are only a few Myr old. Consistently with a young age our
spectra show no evidence for stellar absorption features typical of supergiants, nor of
[FeII] emission typical of supernovae. The K-band gas fraction ranges from 20 to
40 \% showing that gas emission can significantly contaminate broadband near-infrared 
colors in young metal-poor starbursts. We have detected molecular hydrogen in emission in 
all three objects. All sources show bright knots superimposed on a lower surface brighness 
envelope. The knots are identified with Super Star Clusters; six of them
are present in UM~462. In all galaxies we detect the presence of an old stellar population. 
   \end{abstract}

%

\section{Introduction}
Since the discovery of the first two Blue Compact Dwarf (BCD) Galaxies
by Sargent \& Searle (1970) and Searle \& Sargent (1972), and their 
identification by Thuan \& Martin (1981) as a class of low-luminosity
metal-deficient objects, the importance of these galaxies has increased. 
BCD galaxies are characterized by strong episodes of star formation as
evidenced by the blue optical colors and the presence of bright
emission lines in the optical spectra. They 
span a range of metallicities that goes from about a third solar down to 
$Z_{\odot}$/51 for I~Zw~18, indicating a non-uniform star formation history 
possibly undergoing episodic bursts Thuan (1991). 
The importance of these galaxies is obvious since
they are forming stars in an environment that must be similar to that 
expected in primeval galaxies (Izotov \& Thuan 1999). For this reason they 
are unique laboratories to study star formation as it occurred
in the Universe during its earliest phases.

Izotov \& Thuan (1999) have proposed, based on an extensive study of a sample of
BCD galaxies, to use the oxygen abundance as an age indicator; all galaxies with 
12+log(O/H)$\le$ 7.6 would be younger than 40 Gyr. However single star photometry
of one BCD galaxy showed that this limit is actually 1 Gyr (Izotov \& Thuan 2002).
Another approach to the problem of the age is the study of the stellar populations 
based on deep multi-wavelength photometry. 
NIR colors are effective indicators of the age of a stellar population once
the metallicity is known. However
Vanzi et al. (2000) point out how, though the NIR colors of normal, active and 
starburst galaxies are fairly well understood, their interpretation
in BCDs is not easy and certainly not free from ambiguity. This is mainly 
due to two factors,
1) the broadband photometry of BCDs is strongly affected by nebular 
emission and no reliable conclusion on the stellar population can be driven 
without correcting the colors for this effect,
2) the interpretation of the colors is complicated by
the difficultiy of incorporating the effect of metallicity into the
evolutionary models in a satisfactory way. 
The case of SBS~0335-052 (the second lowest metallicity BCD with 
$Z_{\odot}$/40) is a good testbench 
since none of the currently available models is able to reproduce the colors of 
this galaxy even after taking into account the nebular contribution.
To measure the nebular contribution spectroscopy is necessary. NIR spectra
also offer the unique opportunity to probe the molecular hydrogen warm phase and to
study a number of features not accesible at other wavelengths.
Our main motivation for studying BCDs with $Z > Z_{\odot}/10$ is to link the 
well-studied extremely metal deficient BCDs, like I~Zw~18 and SBS~0335-052, with the
dwarf irregulars at the other end of the metallicity range, such as the LMC 
($Z_{\odot}/3$). 
The fact that the properties of sub-solar 
and very low metallicity dwarf galaxies could be smoothly
connected is suggested by the finding of Guseva et al. (2000) that the number of
Wolf-Rayet stars in these galaxies is an increasing function of the metallicity in
agreement with the predictions of the models.
We are particularly interested in the photometric and spectroscopic properties of 
BCDs, their star-formation process, dust, gas and $H_2$ content as a function of
the metallicity.
It is our opinion that this approach can put important constraints on the star formation
history of BCDs and on star formation models.

The present paper is organized as follows. In Section 2 we present our new
observations. Section 3 is devoted to the discussion of the spectra and 
Section 4 to the images. In Section 5 we summarize the conclusions of our work.


\section{Observations}
We have selected Tol~35, Tol~3 and UM~462 as representative of star formation 
at intermediate sub-solar
metallicity. Their abundances are $Z_{\odot}$/6 for Tol~35 and Tol~3
(Kobulnicky et al. 1999) and $Z_{\odot}$/9 for UM~462 (Izotov \& Thuan 1998,
Guseva et al. 2000). All three galaxies display the Wolf-Rayet features at 4686 \AA~ 
(Vacca \& Conti 1992, Schaerer et al. 1999, Guseva et al 2000). This is indicative
of a recent and short star formation episode as shown by the evolution of the number of
WR over O stars predicted by single population evolutionary models (e.g. Leitherer \&
Heckman 1995).
We have calculated distances to the sample galaxies 
using a Hubble constant $H_0$ of 70 km/s/Mpc, and the Virgo
nonlinear flow model defined by Kraan-Korteweg (1986) and Giovanardi (2002).
With this $H_0$ and the model of Kraan-Korteweg, the Virgo distance is 17.0 Mpc
The characteristics of the galaxies are summarized in Table 1.

\begin{table}
\caption{Observational Characteristics of the sample galaxies}
\begin{minipage}{\linewidth}
\begin{tabular}{cccc}
\hline
             &   Tol~35   &  Tol~3     &  UM~462      \\
\hline
R.A. (2000)  &  13:27:06  &  10:06:33  &  11:52:37    \\
DEC. (2000)  & -27:57:24  & -29:56:09  & -02:28:10    \\
v (Km/s)     &  2023      &  865       &  1055        \\
D (Mpc) \footnote{see text for details}
             &   30.3     &  10.2      &  15.3        \\
Z/$Z_{\odot}$&  1/6       &  1/6       &  1/9         \\
m(B)         &  14.5      &  13.5      &  14.5        \\
other names  &Tol~1324-276&Tol~1004-296& Mrk1307      \\
             & IC4249     & NGC~3125   & UGC~6850     \\
\hline
\end{tabular}
\end{minipage}
\end{table}

\begin{table}
\caption{Log of the Observations}
\label{Log}
\begin{tabular}{ccccc}
\hline
object   & grism & $t_{int}$ (min) & PA & aperture(\arcsec) \\
\hline
Tol~35   & blue  & 32 & 124 & 1$\times$3/1$\times$3.5 \\
Tol~35   &  red  & 28 & 124 & 1$\times$3/1$\times$3.5 \\
Tol~3    & blue  & 30 & 113 & 1$\times$1.5/1$\times$3 \\
Tol~3    &  red  & 24 & 113 & 1$\times$1.5/1$\times$3 \\
UM~462   & blue  & 30 &  23 & 1$\times$2 \\
UM~462   &  red  & 24 &  23 & 1$\times$2 \\
\hline
\end{tabular}
\end{table}

\begin{figure*}
\psfig{figure=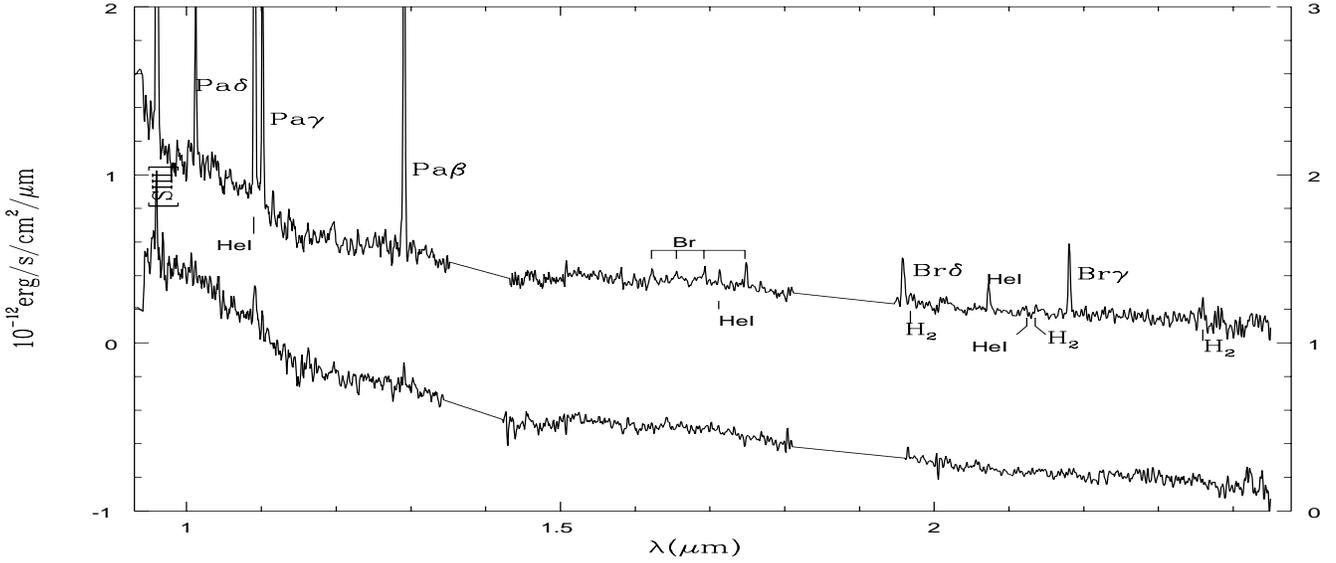,width=18cm,height=8cm,angle=-90}
\caption{NIR spectrum of Tol~35. The lower spectrum is extracted from the 
nucleus - region B (refear to flux in the right scale), the upper one from 
the brightest HII region in the galaxy - region A (refear to flux in the left scale). The regions of poor atmospheric 
transmission have been replaced by straight lines.}
\label{sptol35}
\end{figure*}

\begin{figure*}
\psfig{figure=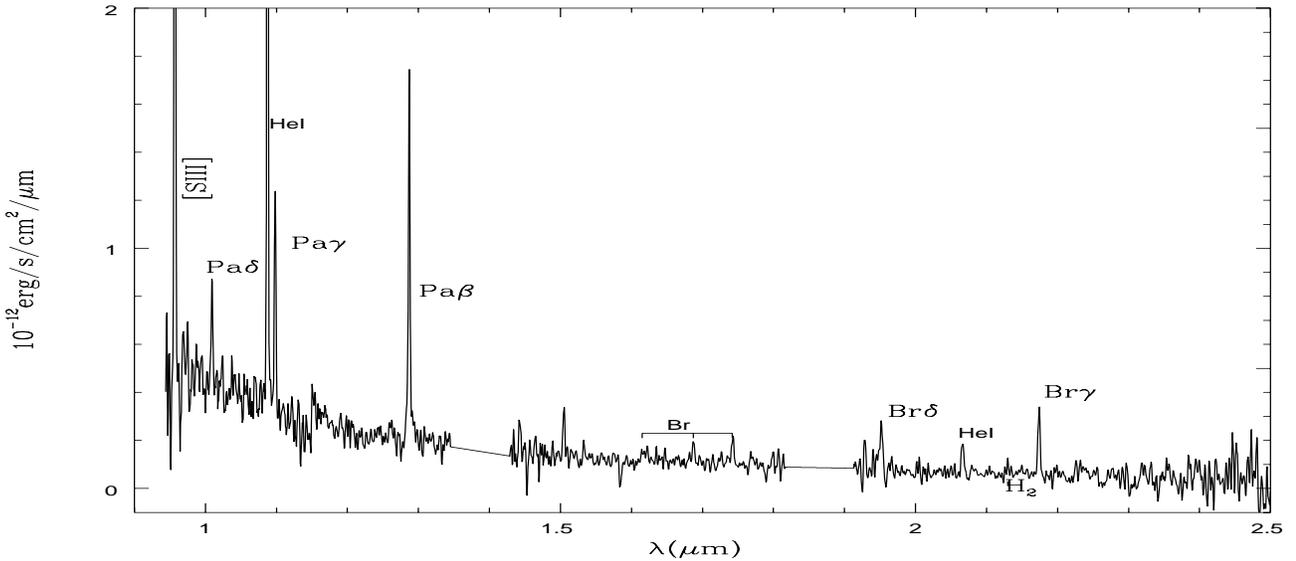,width=18cm,height=8cm,angle=-90}
\caption{NIR spectrum of UM~462 centered on region 1.}
\label{spum462}
\end{figure*}

\begin{figure*}
\psfig{figure=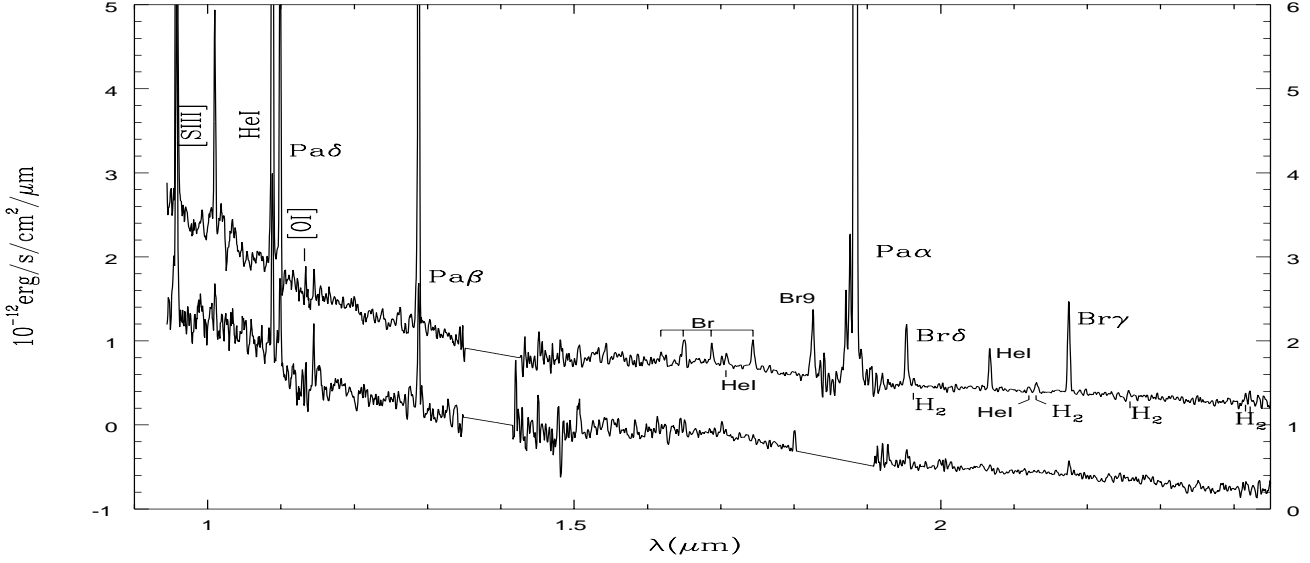,width=18cm,height=8cm,angle=-90}
\caption{NIR spectrum of Tol~3. The lower spectrum is extracted from region B 
(refear to flux in the right scale), the upper spectrum from region A (refear to 
flux in the left scale)}
\label{sptol3}
\end{figure*}

NIR spectra were observed with SOFI in 2000 April and 2001 April
using the low resolution red and blue grisms which give a resolution R=600 with 
a 1\arcsec wide slit. The slit position angle
(PA) was always chosen to include what seemed to be the most interesting
regions in the Ks acquisition image. All
galaxies were observed nodding along the slit. The Log of the observations 
is presented in Table \ref{Log}.
The reduction of the spectra was carried out in IRAF following the standard steps. 1D spectra 
centered on the brightest sources present in the slit 
were extracted with apertures that maximize the S/N, see Table 2. The telluric absorption
features were removed from the spectra dividing by the spectrum of an
O or G type star, then the original spectral shape was reestablished by multiplying
by a black-body of suitable temperature in the first case or by the solar
spectrum in the second one. The spectra were flux calibrated using the 
photometry of broadband images within the spectra extraction apertures. 
The agreement between the slope of the spectra and the photometry was at
the 10\% level for Tol~3 and Tol~35 and at the 20\% level for UM~462. 
For this reason
a correction factor was applied to the spectra. The comparison 
between the spectral fluxes and the photometry was always
done by integrating the spectrum over the filters bandpass. Since the atmosphere
significantly affects transmission, especially in the J filter its effect was included
as well. The flux-calibrated
spectra are displayed in Fig. \ref{sptol35}, \ref{spum462} and \ref{sptol3}. 
The detected emission lines are listed in Tab. \ref{lines}.


\begin{table*}
\caption{NIR Emission lines, fluxes measured in 10$^{-15}$\,erg/s/cm$^2$}
\label{lines}
\begin{tabular}{cccccc}
\hline
line        & $\lambda_{rest}$ & Tol~35~A & Tol~3~A & Tol~3~B & UM~462 \\
\hline
[SIII]+Pa 8 & 0.953  & 23.5$\pm$0.5 & 48.2$\pm$0.3 & 12.8$\pm$0.3 &  8.7$\pm$0.3 \\  
Pa$\delta$  & 1.005  &  3.1$\pm$0.2 &  6.6$\pm$0.2 &  5.8$\pm$0.3 &  1.2$\pm$0.3 \\
HeI         & 1.082  & 15.5$\pm$0.3 & 36.1$\pm$0.3 &  2.5$\pm$0.3 & 10.5$\pm$0.5 \\
Pa$\gamma$  & 1.093  &  4.1$\pm$0.2 & 11.8$\pm$0.3 &  1.7$\pm$0.3 &  2.3$\pm$0.3 \\
$[OI]$      & 1.128  &       -      &  0.5$\pm$0.2 &      -       &       -      \\
Pa$\beta$   & 1.282  &  8.3$\pm$0.2 & 18.2$\pm$0.2 &  4.1$\pm$0.2 &  4.6$\pm$0.3 \\
  ?         & 1.498  &  0.2$\pm$0.1 &      -       &      -       &  0.6$\pm$0.2 \\
Br13        & 1.611  &  0.2$\pm$0.1 &      -       &      -       &       -      \\ 
Br12        & 1.641  &  0.2$\pm$0.1 &  1.2$\pm$0.3 &      -       &       -      \\ 
Br11        & 1.681  &  0.4$\pm$0.1 &  0.8$\pm$0.1 &      -       &  0.3$\pm$0.1 \\
HeI         & 1.701  &  0.2$\pm$0.1 &  0.5$\pm$0.1 &      -       &       -      \\
Br10        & 1.736  &  0.5$\pm$0.1 &  1.2$\pm$0.1 &      -       &       -      \\
Br9         &        &  3.7$\pm$?   &  0.4$\pm$0.1 &      -       &       -      \\
Br$\delta$  & 1.944  &  1.3$\pm$0.1 &  2.8$\pm$0.1 &  0.7$\pm$0.3 &  1.1$\pm$0.1 \\
H$_2$(1-0)S(3) & 1.957  &  0.2$\pm$0.1 &  0.2$\pm$0.1 &       -       &       -     \\
HeI         & 2.058  &  0.6$\pm$0.1 &  1.8$\pm$0.1 &      -       &  0.6$\pm$0.1 \\
HeI         & 2.113  &       -      &  0.3$\pm$0.1 &      -       &       -      \\
H$_2$       & 2.121  &  0.2$\pm$0.1 &  0.6$\pm$0.1 &      -       &  0.1$\pm$0.1 \\
Br$\gamma$  & 2.165  &  1.6$\pm$0.1 &  4.1$\pm$0.1 &  0.6$\pm$0.2 &  1.1$\pm$0.1 \\
H$_2$       & 2.248  &       -      &  0.2$\pm$0.1 &      -       &       -      \\
H$_2$       & 2.407  &       -      &  0.4$\pm$0.2 &      -       &       -      \\
H$_2$       & 2.413  &       -      &  0.4$\pm$0.2 &      -       &       -      \\
\hline
\end{tabular}
\end{table*}

Broadband images in the J, H and Ks NIR filters were acquired with
SOFI at the ESO-NTT in 2000 July. 
The integration time was 10 minutes per filter for
all galaxies. All galaxies being small compared to the field of view 
they were observed dithering ON source. The images were reduced 
using the ESO Eclipse Package. The accuracy of the photometry was always 
better than 3\%. Images of the three galaxies in the Ks band
are shown in Fig. \ref{imtol35}, \ref{imtol3} and \ref{imum462}.

\section{NIR Spectra}
\begin{figure}
\psfig{figure=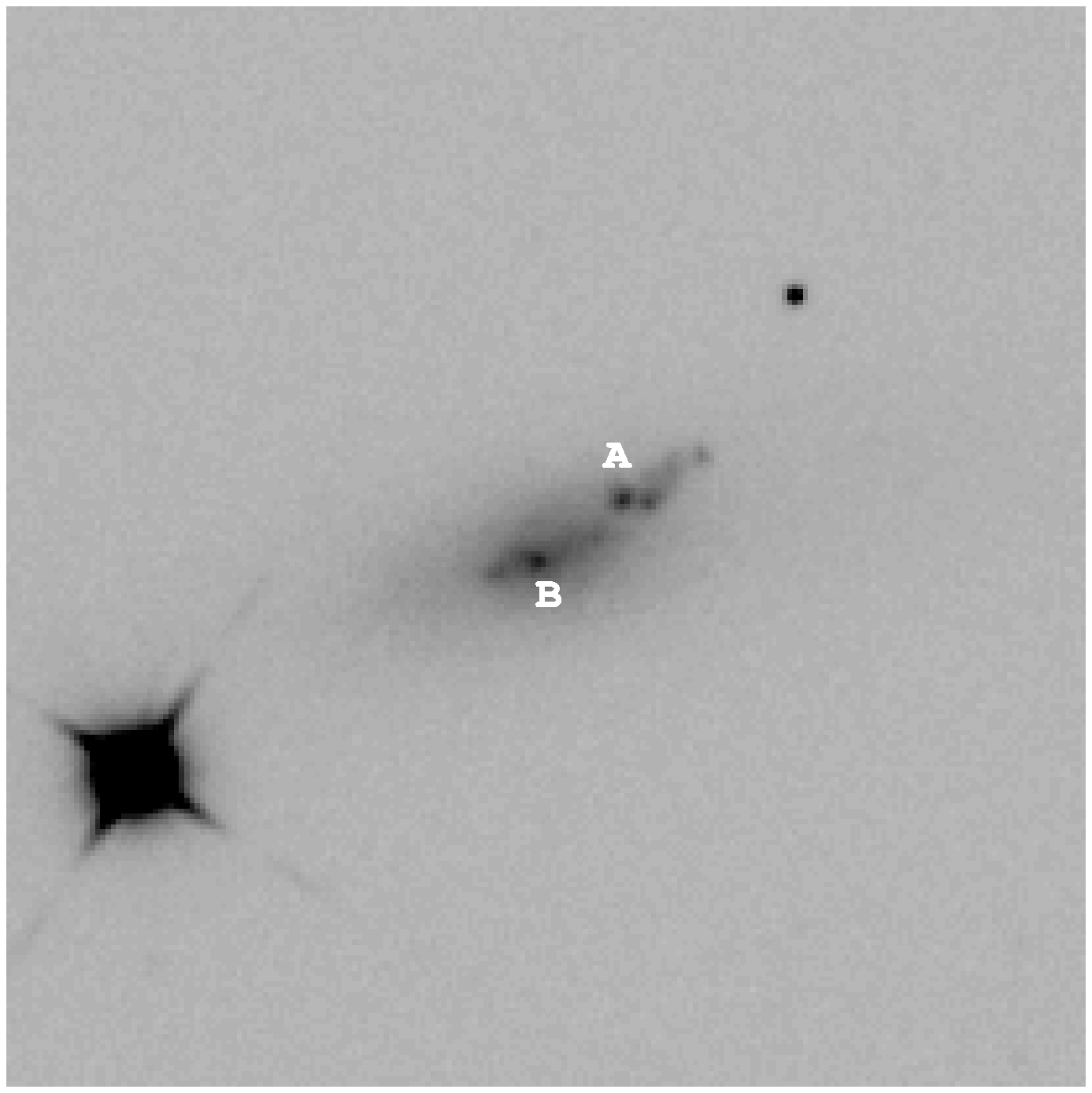,width=8cm,angle=0}
\caption{Image of Tol~35 in $K_S$. The field is 1$\times$1 arcmin, north is up est at
left. A and B mark the sources for which we extracted spectra.}
\label{imtol35}
\end{figure}

\begin{figure}
\psfig{figure=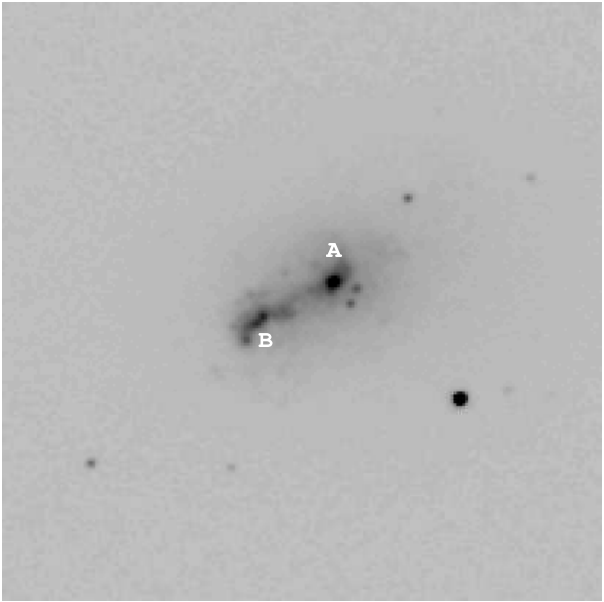,width=8cm,angle=0}
\caption{Image of Tol~3 in $K_S$. A and B mark the sources for which we extracted spectra. Field and orientation as in the previous images.}
\label{imtol3}
\end{figure}

\begin{figure}
\psfig{figure=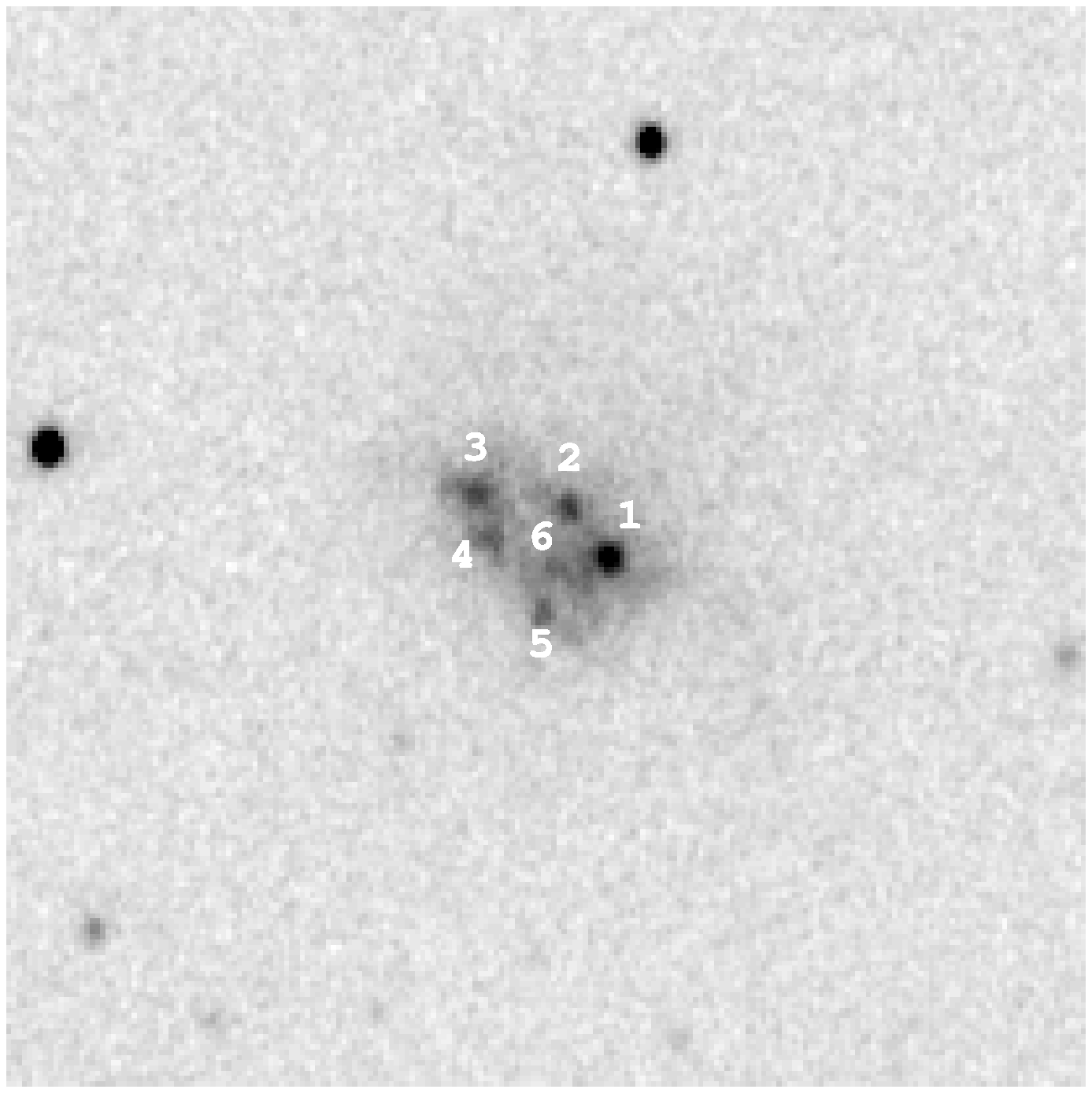,width=8cm,angle=0}
\caption{Image of UM~462 in $K_S$. Numbers from 1 to 6 mark the regions for which we 
extracted photometry. The spectrum is centered on region 1. Field and orientation as 
in the previous image.}
\label{imum462}
\end{figure}

In all three galaxies we have observed HII-region-like spectra with clear signatures of
powerful episodes of star formation. These are the regions A in Tol~35 and Tol~3
(see Fig. \ref{imtol35} and \ref{imtol3}) and region 1 in UM~462 (see Fig. \ref{imum462}).
In particular, bright recombination lines of HI and
HeI are detected in all the spectra. 


\subsection{Extinction and Star Formation Rates}
From the ratio of $Pa\beta / Br\gamma$ it is possible to calculate the extinction
toward the observed regions. We assume the intrinsic value 5.86 for the ratio as
given by Hummer \&
Storey (1987) for $T=10^4$K and $n=10^2~cm^{-3}$ and the extinction law of Rieke \&
Lebofsky (1985). Galactic extinction is not negligible toward Tol~35 and Tol~3,
$A_V=$0.270 and 0.327 respectively, so that we have corrected for this effect first.
Though the errors are large, a
non-negligible extinction is observed in all cases. We obtain $A_V=0.7\pm0.4$ in Tol~35~A,
$A_V=1.6\pm0.2$ in Tol~3~A, $A_V=3.3\pm2.3$ in Tol~3~B and $A_V=2.6\pm1.8$ in UM~462~1.
These estimates are in good agreement with the optical values for the first two galaxies. 
Vacca \& Conti (1992) give
$A_V=0.77$ for Tol~35~A and 1.97 for Tol~3~A. From the data of Terlevich et al. 
(1991) we derive $A_V=0.26$ for UM~462 while Guseva et. al (2000) measure $A_V=0.27$.
Although the discrepancy in the case of UM~462 could be possibly due
to a mis-match in slit position (the morphology of UM~462 seems in fact
quite different in the optical and NIR - see Section 4), 
such a large difference could also be ascribed to the presence of hidden star formation
somewhat similar to, but less extreme than in SBS~0335-052 (Hunt et al. 2001). Dust could 
be well present in UM~462 that has been detected by IRAS at 60 and 100 $\mu m$.
We calculated the star formation rate in this galaxy using the global $H\alpha$ flux 
from Guseva et al. (2000) ($8.17~10^{-12}~erg/s/cm^2$) and the IRAS fluxes 
($F_{60}=0.944$ and $F_{100}=0.896$ Jy) with the conversion factors of Kennicutt (1998)
obtaining $SFR(H\alpha)=1.4~M_{\odot}/yr$ and $SFR(IR)=0.06~M_{\odot}/yr$. 
From this comparison we do not find strong evidence to support the presence of 
hidden star formation so that the issue must be investigated further.

Based on our $Br\gamma$ fluxes corrected for the extinction we measured a SFR of
0.28, 0.24 and 0.24 $M_{\odot}/yr$ respectively in Tol~35~A, Tol~3~A and UM~462 or
$4.4~10^{-3}$, $3.3~10^{-2}$ and $2.2~10{-2}~M_{\odot}/yr/Kpc^2$.

\subsection{Molecular Hydrogen}
Molecular hydrogen emission lines are detected in all galaxies. 
The (1,0)S(1)/Br$\gamma$ ratios are 0.15, 0.12 and 0.09 respectively 
for Tol~3~A, Tol~35~A and UM~462. They are
similar to the value measured in SBS~0335-520 by Vanzi et al.
(2000) and perfectly consistent with the trend reported by Vanzi \& Rieke (1997), 
suggesting that the $H_2$ emission is not affected, or it is 
affected at a low level, by the metallicity. The fact that the $H_2$2.12 flux is
only marginally affected by the metallicity in BCD galaxies is a remarkable finding:
it tells us that $H_2$ is probably clumpy, confined to the star forming regions
dust-enriched by SNe and not diffuse, consistently with the observations of FUSE 
(e.g. Vidal-Madjar et al. 2000, Thuan et al. 2002).
The detection of the (1,0)S(3) transition in Tol~35 and Tol~3 is not significant
enough to derive
any conclusion on the excitation mechanism but the reddest part of the spectrum of Tol~3
is good enough to allow unambiguous detection of the transitions (2,1)S(1)2.248, 
(1,0)Q(1)2.407 and (1,0)Q(2)2.413. These detections strongly support the fluorescent
excitation of $H_2$ as can be easily seen from the models reviewed by Engelbracht et al.
(1998). Such a mechanism is quite natural in regions dominated by a strong UV
field from young massive stars. From the similarity of the objects under study
we can assume that the same mechanism is at work also in Tol~35 and UM~462 and
explain the low (1,0)S(1)/Br$\gamma$ ratio observed as produced by large "nude" 
clusters of young stars (Vanzi et al. 2000, Puxley et al. 1990).

\subsection{[FeII] lines}
The NIR lines of [FeII] are often used as indicators of supernovae in starburst
galaxies (e.g. Vanzi \& Rieke 1997) since the ratio [FeII]/Br$\gamma$ has been found
high ($>20$) in galactic SN remnants and low ($<0.1$) in galactic HII regions 
(Moorwood \& Oliva 1988).
[FeII] is not detected in any of the regions observed, either at 1.26 or at 1.64 $\mu m$, 
meaning that the spectra are dominated by young episodes of star formation 
where even the most massive stars have not yet had time to evolve to become SN.
In other words the star formation observed in Tol~35~A, Tol~3~A and UM~462 must
be younger than about 10 Myr. This conclusion is also supported by the very
high equivalent width of $Br\gamma$, 90, 110 and 170 \AA~ respectively in the 3
galaxies. According to Starburst99 (SB99, Leitherer et al. 1999), using an 
instantaneous burst, $Z_{\odot}/5$ and a Salpeter IMF, these correspond to 
ages between 5 and 7 Myr.
Further support to the extreme youth of the observed episodes of
star formation is given by the HeI lines.
The HeI line at 1.700 $\mu m$ is clearly detected in region A of Tol~3 and 
region A of Tol~35. The HeI line at 2.113 $\mu m$ is detected in Tol~3
only. Both lines are strong signatures of the presence of young 
massive stars as discussed by Vanzi et al. (1996).
The ratios HeI1.7/Br10 and HeI2.11/Br$\gamma$ in Tol~3~A, respectively 0.42 
and 0.07, and the ratio HeI1.7/Br10=0.40 in Tol~35~A, are in good agreement 
with the saturated values calculated by Vanzi et al. (1996).
The two lines are not detected in UM~462 whose spectrum is however of lower quality.
The presence of Wolf-Rayet stars in all three BCDs also gives an age of about 5-6
Myr for the young massive stellar population (Guseva et al. 2000).

\subsection{Other Spectral Features}
The [SIII] line at 0.953 $\mu m$ is detected in all the spectra. Diaz et al.
(1985) use this line as a diagnostic of the excitation mechanism. We took the
ratio [SIII]/$Pa\beta$, after subtracting the contribution of Pa8 that is 
blended with [SIII] at our resolution, and converted 
it to [SIII]/$H\alpha$ using a standard value for $H\alpha/P\beta$. Our values 
lie at the extreme high-excitation end of the HII region area in the plot of 
Diaz et al. (1985).
This is easily explained by the low metallicity of our galaxies compared to the
galactic HII regions used by other authors. The [SIII] line is in fact sensitive to
the sulfur abundance (Rudy et al. 2001). 

The CO absorption band at 2.29\,$\mu$m can be used to constrain the age of 
the star-formation episode in starburst galaxies (Doyon et al. 1994), although it is not
particularly reliable for ages older than roughly 10\,Myr (Origlia \& Oliva 2000), 
when red supergiants begin to emerge.
In low metallicity environments, CO absorption is generally weak mostly 
because carbon abundance is lower, but also because stellar temperatures are warmer 
than those in more metallic systems (Origlia et al. 1997). Our NIR spectra 
reveal no absorption bands either at the CO(6,3) bandhead at 1.62\,$\mu$m or at 
2.29\,$\mu$m, the CO(2,0) bandhead. In addition to the metallicity effect, 
this non-detection can be explained also by the young age of the bursts.

\section{NIR images}
\subsection{Morphology and colors}
As seen in the images presented in Figs. \ref{imtol35}, \ref{imtol3}, and \ref{imum462},
all galaxies show two or more bright knots of intense star-formation activity, surrounded 
by low-surface brightness envelopes. Because of the non-ellipticity of its outer isophotes,
UM\,462 would be classified as iI according to the classification scheme of Loose \& Thuan 
(1986). Tol\,3 and Tol\,35, on the other hand, present rather regular elliptical
isophotes in the outer regions (although the foreground star
disturbs the appearance toward the SE in Tol\,35), and therefore would
be classified as iE, the most common morphology of BCDs.
In terms of morphology, Tol\,3 and Tol\,35 are reminiscent of I\,Zw\,18, which is dominated
by two main bright centers of star formation surrounded by a low
surface-brightness envelope.
UM\,462 shows a peculiar morphology characterized by several
knots of varying brightness, embedded within an irregular outer envelope.
However a deeper B optical image of UM\,462 by Cair\'os et al. (2001)
does show that the outermost contours do become elliptical, and thus should be
classified as iE.
In all galaxies, the NIR colors and spectra of the knots differ, and
may represent examples of propagating star formation
(see Section \ref{sscs}).

\begin{figure}
\hspace{-1.0cm}
\hbox{
\psfig{figure=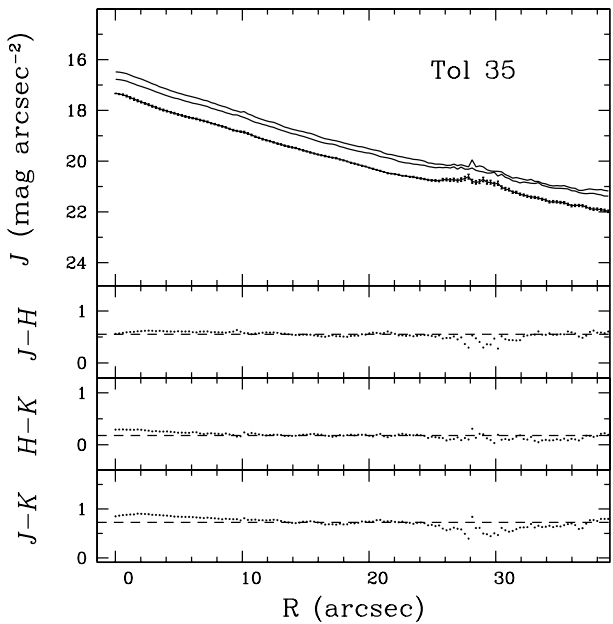,width=9.5cm}
}
\caption{Surface-brightness profiles of Tol\,3 (left panel) and Tol\,35 (right panel).
In the upper panels, the $J$ band profile is shown by data points with 
uncertainties, together with the $H$ and $Ks$ bands shown by solid lines.
The lower panels show the radial variation of the NIR colors, with the mean 
colors of the outer isophotes shown as horizontal dashed lines.
The surface-brightness profile of Tol\,35 is compromised
\label{fig:profiles-a}
}
\end{figure}

To better constrain the colors of the extended regions,
we have derived surface-brightness profiles of the galaxies.
The profiles were extracted by first fitting elliptical isophotes
to the $J$-band image,
In Tol\,3 and Tol\,35, ellipse centers were fixed to the brightest
knot (Tol\,3A and Tol\,35B);
in the case of UM\,462, the ellipse center was assigned to be
the center of symmetry of the outer isophotes.
In this case, the surface brightness peaks integrated azimuthally
appear as a ``shoulder'' in the profile at $R\,\sim\,$3\,\arcsec.
While average-sigma clipping was performed for all objects,
for Tol\,35 it was also necessary to apply a mask in order to
minimize the effect of the bright foreground star;
nevertheless, the effects of the star appear in the profile at
$R\,>\,$28\,\arcsec.
We note also that Tol~35B corresponds to the center of symmetry of
the outer isophotes, and is most probably the nucleus of the
galaxy.
The surface-brightness profiles are described by exponential laws
in Tol~35 and Tol~3 and by a higher order generalized exponential in
UM~462.
Mean colors of the outer regions of the galaxies were estimated
from the profiles by averaging; these are shown by horizontal
dashed lines in the lower panels of Figs. \ref{fig:profiles-a}
and \ref{fig:profiles-b} and reported in Table \ref{colors}.
We have calculated the colors in the spectroscopic apertures,
and performed photometry of the bright knots.
All colors have been corrected for Galactic extinction (Schlegel et al. 1998).
For the knots, we used a photometric aperture roughly equal to
the seeing width (1\arcsec),
appropriate for characterizing colors of point sources
superimposed on a variable background.
In Table \ref{colors} we list the colors observed in the spectroscopic apertures,
and the corrections for nebular emission.
The continuum correction was estimated by using the coefficients in Joy
\& Lester (1988) and our Br$\gamma$ fluxes.
The correction for emission lines was calculated by summing over
the lines in our spectra.
Total gas fractions are also reported in Table \ref{colors}, and range from 20 --
40\% in the brightest knots; $J$ band fractions tend to be slightly
larger than those in $Ks$ because of the strong emission lines around
1\,$\mu$m.
Such large gas fractions and the corresponding color corrections
clearly illustrate how ionized gas significantly affects broadband
colors of young HII regions.
\subsection{Models}
 
\begin{figure*}
\flushleft{\psfig{figure=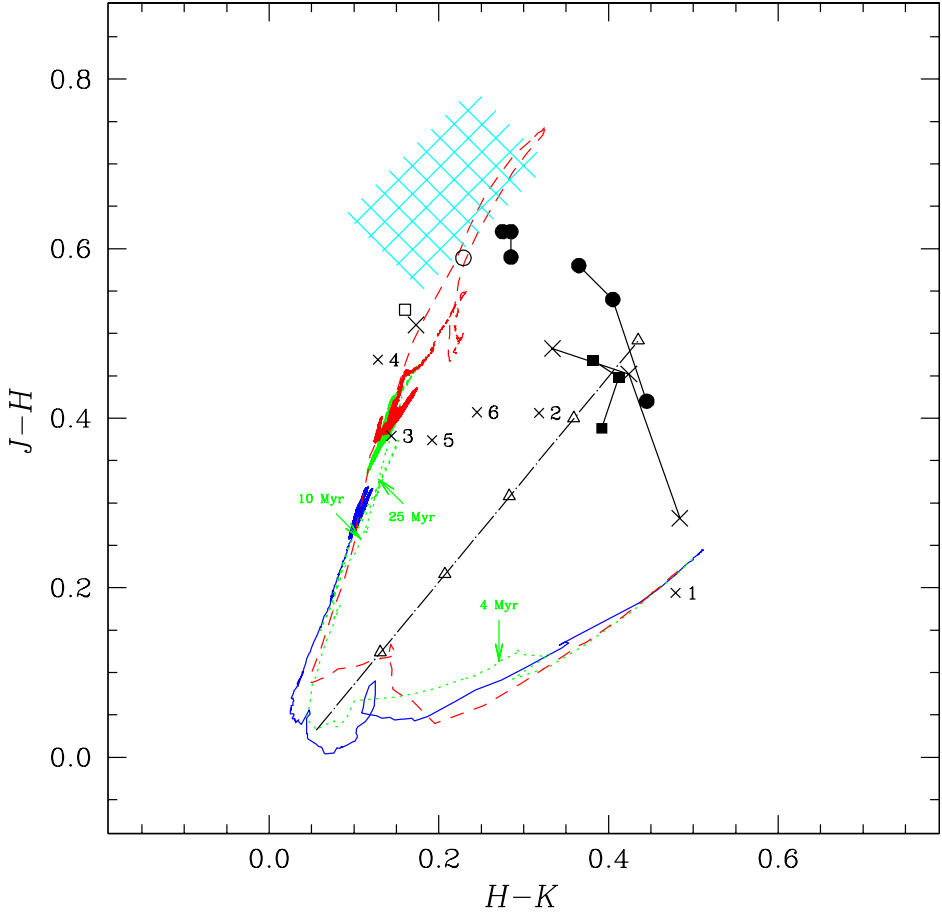,width=14.0cm}}
\caption{$J-H$ vs. $H-K$ diagram for bright knots and extended regions of
sample galaxies.  Tol~3 is represented by circles, Tol~35
by squares, and UM~462 by crosses (with number of knot when applicable).
Filled symbols show photometry for the spectroscopic aperture, and
open ones the mean colors of the outer regions. Colors observed and corrected
for the gas contribution are plotted and connected by solid lines.
SB99 models are shown for 
three metallicities: 1/20 (blue, solid line), 1/5 (green, dotted line), 
and solar (red, dashed line). The cyan-colored grid rapresents the empirical NIR colors of
late-type and dwarf galaxies. 
An extinction vector starts from the point of age 7 Myr on the $Z_{\odot}/5$ model,
points corresponding to $A_V=1$ are marked by open triangles.
\label{fig:colors}
}
\end{figure*}

\begin{table*}
\caption{NIR magnitudes and colors observed on the spectroscopic apertures, gas fractions,
corrections for the contribution of the ionized gas and colors of the extended regions of
the galaxies.}
\label{colors}
\begin{tabular}{cccccccccc}
\hline
Name  &   & J  & J-H & H-K & $(F_{gas}/F_{tot})_J$ & $(F_{gas}/F_{tot})_H$ & $(F_{gas}/F_{tot})_K$ & $\Delta (J-H)$ & $\Delta (H-K)$ \\
\hline
Tol3  & A & 15.89 & 0.45 & 0.47 & 0.30 & 0.19 & 0.24 & 0.16 & -0.08 \\
      & B & 15.91 & 0.62 & 0.31 & 0.05 & 0.02 & 0.03 & 0.03 & -0.01 \\ 
   & ext. &       & 0.60 & 0.25 &      &      &      &   &   \\
\hline
Tol35 & A & 16.67 & 0.41 & 0.41 & 0.25 & 0.19 & 0.20 & 0.08 & -0.01 \\
      & B & 16.50 & 0.60 & 0.27 & 0.00 & 0.00 & 0.00 & 0.00 &  0.00 \\
   & ext. &       & 0.55 & 0.18 &      &      &      &   &   \\
\hline
UM462 &   & 17.74 & 0.29 & 0.49 & 0.42 & 0.30 & 0.39 & 0.20 & -0.15 \\
   & ext. &       & 0.53 & 0.19 &      &      &      &   &   \\
\hline
\end{tabular}
\end{table*}

In Figure \ref{fig:colors}, we compare our results with models: Tol~3 is represented 
by circles, Tol~35 by squares, and UM~462 by X's (with number of knot when applicable).
Filled symbols show photometry for the spectroscopic aperture, and
open ones the mean colors of the outer regions.  For the spectral
apertures, photometry corrected for gas emission is shown (first lines,
then continuum$+$lines).
SB99 models are shown for three metallicities: 1/20 (blue, solid line),
1/5 (green, dotted line), and solar (red, dashed line) for a Salpeter IMF
with upper cut-off at $100~M_{\odot}$.
For 1/5 $Z_\odot$, ages of 4, 10, and 25\,Myr are marked with arrows.
The characteristic shape of the curves is given by the fact that
young ages are characterized by the reddest $H-K$ colors, because of the 
dominant gas (continuum only in SB99) emission such emission can dominate
the K band (see Vanzi et al. 2000). Empirical NIR colors of 
late-type and dwarf galaxies (de Jong 1996) are shown as a cyan-colored grid.
The X's numbered from 1 to 6 correspond to the knots detected in
UM~462, and will be discussed below.

As in the majority of BCDs,
the colors of the extended emission are in quite good agreement
with evolved stellar populations of 1/5 $Z_\odot$ to $Z_\odot$ metallicity.
It is clear that in each of our sample galaxies, at metallicities
from $Z_\odot/9$ to $Z_\odot/6$, the extended envelope is populated
by evolved (age $>$\,5\,Gyr) stars.
Judging from $J-H$, Tol~3 is the most evolved galaxy; the knot B colors
are very similar to those of the extended envelope, and both are
relatively red in $J-H$ and blue in $H-K$.
 
To compare the colors of the knots with the models,
it is necessary to correct the observations for gas line emission,
since SB99 includes the contribution of the nebular continuum only.
The only knot clearly compatible with the models is knot 1 in UM~462,
although the colors have not been corrected for line emission.
The remaining knots in Tol~3 (circles), Tol~35 (squares), and UM~462
(X's) are roughly compatible with a stellar population younger than 10\,Myr,
and a few magnitudes of extinction.
Such young ages are consistent with the Br$\gamma$ equivalent widths
and the presence of HeI emission lines in our spectra;
they are also consistent with the presence of Wolf-Rayet stars in these
galaxies (Guseva et al. 2000, Vacca \& Conti 1992).
 
The brightest knots in each galaxy show a peak red
$H-K\,=\,0.5$, in contrast to the surrounding $H-K$ color of 0.2.
Such a red color is an almost certain signature of gas or hot dust,
especially since we see
no spectral sign of red supergiants but do detect large extinction and
large gas fraction.
The brightest knots also show a blue $J-H$ color:
while the surrounding regions and the extended envelopes have
$J-H\,=\,$0.5--0.6, typical of evolved stellar populations,
the bright knots have $J-H\,=0.2$, consistent
with young stars$+$gas (the $J-H$ color of gas including lines
is $\sim$\,0.0--0.2).
Such colors are indicative of extreme youth, since they are
almost certainly due to a high gas fraction plus perhaps a small
amount of hot dust that reddens $H-Ks$.

While the data and the models appear to be in moderate disagreement,
it should be borne in mind that
the corrections for nebular emission and extinction are uncertain.
Moreover, the corrected colors tend to be slightly redder than the models
in $H-K$, which could point to the presence of hot dust
(Vanzi et al. 2000).
Indeed, 500~K dust does not affect $J-H$, but reddens $H-K$,
making this color a strong diagnostic for hot dust emission.

\subsection{Young Compact Star Clusters \label{sscs}}
Table \ref{ssc} lists the photometric properties of the bright knots
in the sample galaxies.
All of the bright knots have absolute $Ks$ magnitudes
ranging from
$-12.5$ in the weakest knot (\#4 in UM~462) to $-15.6$ in Tol~35A;
Tol~35B, the probable galactic nucleus, has $M_{Ks}\,=\,-15.8$,
the brightest knot in our sample.
These have not been corrected for gas emission, so are essentially
upper limits; cluster \#1 in UM~462 with a $Ks$ gas fraction of $\sim$\,40\%,
would be diminished by roughly 0.6\,mag.
 
We have measured the diameter of the clusters, taking into account
the resolution limit dictated by our seeing
(ranging from 0.8 -- 1.1\arcsec\ in $Ks$);
1\arcsec\ corresponds to 50, 147, and 74 pc for
Tol\,3, Tol\,35, and UM\,462, respectively.
Virtually all of the clusters are barely resolved or unresolved,
having diameters ranging from $\sim\,$40\,pc in UM\,462 and
Tol\,3 (the nearest galaxies) to $\sim\,$70\,pc or larger in Tol\,35
(the most distant).
These are almost certainly size upper limits because of the seeing
and distance constraints.

\begin{table}
\caption{SSC properties}
\label{ssc}
\begin{tabular}{ccccccc}
\hline
Name  &   Knot  &   J-H  &  H-K   &  Ks(1") &M$_{Ks}$& Size (pc) \\
\hline
Tol3  &   A     &   0.47 &  0.46  &  15.32  &  -14.72 &    37  \\
      &   B     &   0.64 &  0.29  &  16.13  &  -13.91 &    38  \\
      &   B-tip &   0.27 &  0.47  &    --   &    --   &  \\
\hline
Tol35 &   A     &   0.25 &  0.49  &  16.79  &  -15.61 &    72  \\
      &   B     &   0.56 &  0.28  &  16.63  &  -15.78 &   173  \\
\hline
UM462 &   1     &   0.19 &  0.48  &  17.61  &  -13.31 &    42  \\
      &   2     &   0.41 &  0.32  &  18.14  &  -12.79 &    55  \\
      &   3     &   0.38 &  0.14  &  18.20  &  -12.73 &    86  \\
      &   4     &   0.47 &  0.13  &  18.38  &  -12.55 &    55  \\
      &   5     &   0.37 &  0.19  &  18.45  &  -12.48 &    66  \\
      &   6     &   0.41 &  0.25  &  18.45  &  -12.47 &  $\sim$\,100\\
\hline
\end{tabular}
\end{table}

To derive the number of equivalent O7V stars in the regions
observed spectroscopically and listed in Table \ref{colors}, we
first correct the Br$\gamma$ fluxes for extinction, then multiply
by 35.94 to convert to H$\beta$ fluxes
(for 100 cm$^{-3}$ and T = 10000 K, Osterbrock 1989). We
then convert the H$\beta$ fluxes to the number of Lyman
continuum photons using the conversion factor given by Guseva et al. (2000).
Adopting 10$^{49}$ s$^{-1}$ as the number of Lyman continuum photons
for an O7V star, we obtain 462 O7V stars in Tol3 A,
1435 O7V stars in Tol 35 A, 695 O7V stars in Tol3 B, and 307 O7V stars in
UM462-1. 
The absolute magnitudes, size and number of stars of the clusters
observed are indicative of Super Star Clusters (SSCs) (Billet et al. 2002,
Whitmore 2000).

The SSCs in UM\,462 vary in NIR color, and appear to follow
a trend with age or gas fraction or both.
Knot \# 1 has the reddest $H-Ks$ and bluest $J-H$ color, followed by the two
nearest knots (\#2 and \#6).
While $H-Ks$ gets bluer with increasing knot number (and distance from
knot \#1),
$J-H$ remains roughly constant at 0.4 also for knots \#3 and \#5,
but gets redder for knot \#4.
Such a trend suggests a decreasing gas fraction (because of the decreasing
$H-Ks$ color) going from knot \#1, \#2, \#6, to knot \#4.
This may also be interpreted as an age trend, since gas fraction
decreases with age, and $J-H$ for knot \#4 is slightly redder.
Without spectral information for the different knots, it is difficult
to be more definite.
Nevertheless, the colors of the knots in UM\,462 suggests that the
star formation in knot is \#1 is more recent than that in the remaining
knots; knot \#4 appears to be the oldest.
The knots in Tol\,3 show similar trends, with knot B appearing to be
older than knot A. This is suggestive of propagating star formation,
with the shock waves triggered by supernovae in the older knots
triggering star formation in the younger knots.

\begin{figure}
\vspace{-1.2cm}
\psfig{figure=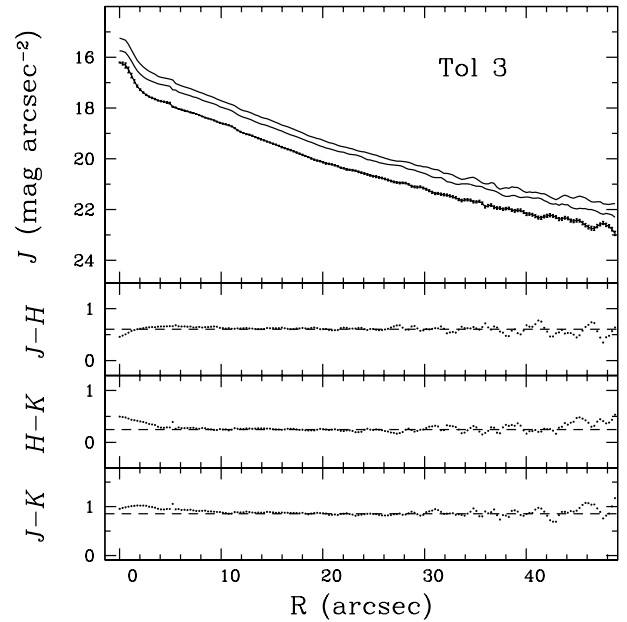,width=9.5cm}
\caption{Surface-brightness profiles of Tol~3. Same as previous Figure.
\label{fig:profiles-b}
}
\end{figure}
\begin{figure}
\vspace{-1.2cm}
\psfig{figure=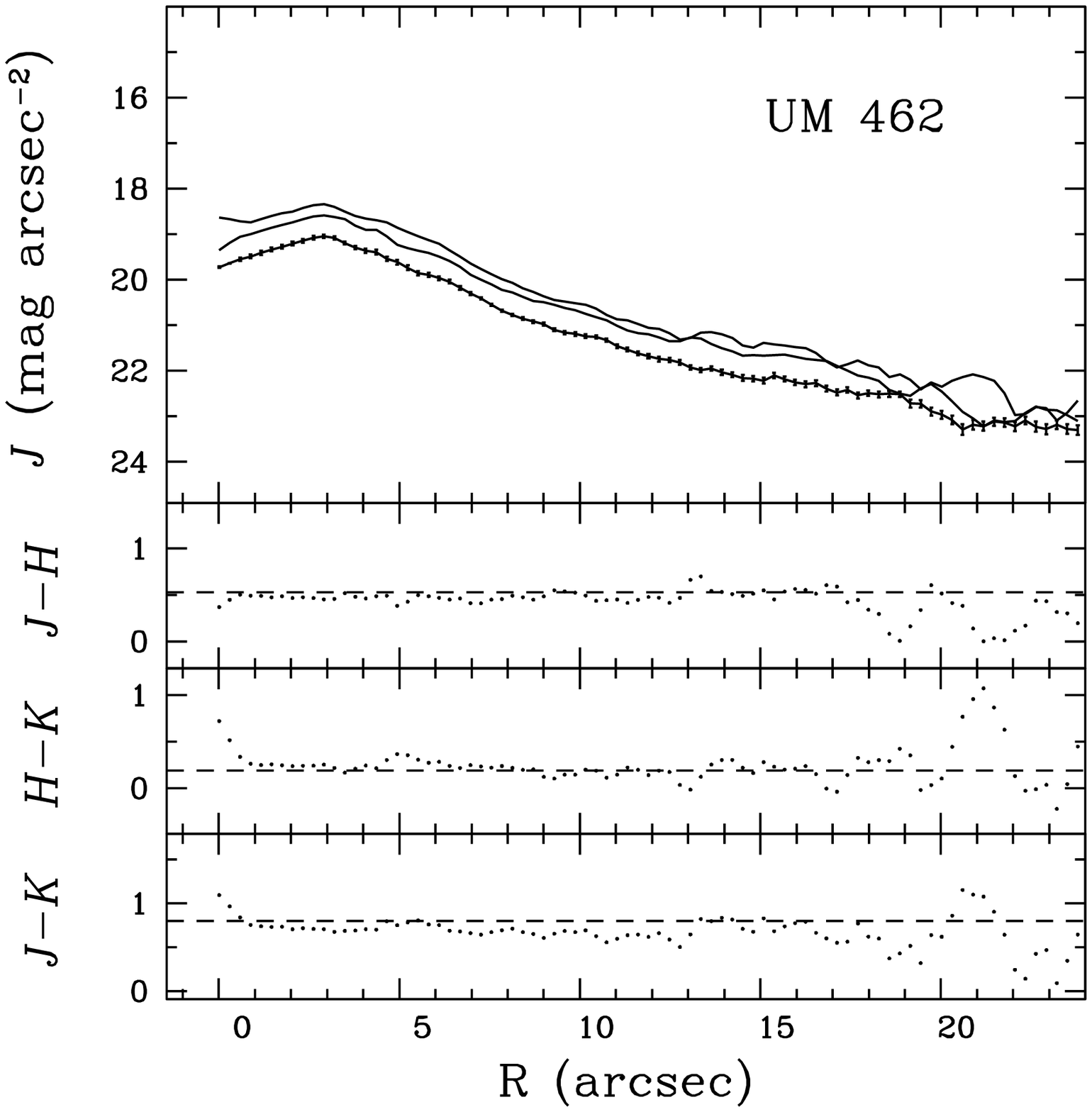,width=9.5cm}
\caption{Surface-brightness profiles of UM\,462. Same as previous Figure.
\label{fig:profiles-b}
}
\end{figure}

\section{Conclusions}
The main results of this work are the following:
   \begin{enumerate}
      \item We have obtained NIR spectra and images of Tol~35, Tol~3 and UM~462 and 
detected bright HII regions in all them. The star-formation episodes are very
young in the brightest regions, not older than few Myr.
      \item Molecular hydrogen lines are present in all galaxies, and trends with
metallicity suggest that in BCDs $H_2$ is clumpy, rather than diffuse.
      \item $Ks$ band gas fractions range from 3 to 40 \%, implying that ionized gas
significantly affects broadband colors.
      \item In all galaxies we detect the presence of Super Star Clusters. There are six
of them in UM~462, arranged in an age or decreasing gas fraction sequence. The SSCs
contain several hundred to more than 1000 O7V stars.
      \item The low-surface brightness envelopes of all galaxies have NIR colors 
representative of evolved star. 
   \end{enumerate}

\begin{acknowledgements}
We are grateful to Jason Spyromilio for generously sharing with us part of
his observing time.
\end{acknowledgements}

\end{document}